\documentclass{article}
\usepackage[T1]{fontenc} 
\usepackage[utf8]{inputenc} 
\usepackage{ismir,amsmath,cite,url}
\usepackage{graphicx}
\usepackage{color}
\def\lbd{}	

\usepackage{lineno}

\title{TOTAL VARIATION IN POPULAR RAP  VOCALS from 2009-2023: extension of the analysis by Georgieva, Ripollés, \& McFee}

\fourauthors
  {Kelvin Walls} {Independent Researcher \\ {\tt kelwalls@gmail.com}}  
  {Iran R. Roman} {New York University\\ {\tt roman@nyu.edu}}
  {Bea Steers} {New York University\\ {\tt bsteers@nyu.edu}}
  {Elena Georgieva} {New York University\\ {\tt elena@nyu.edu}}

\def\authorname{K. Walls, IR. Roman, B. Steers, E. Georgieva}

\begin{document}

\maketitle
\begin{abstract}
Pitch variability in rap vocals is overlooked in favor of the genre's uniquely dynamic rhythmic properties. 
We present an analysis of fundamental frequency (F0) variation in rap vocals over the past 14 years, focusing on song examples that represent the state of modern rap music. 
Our analysis aims at identifying meaningful trends over time, and is in turn a continuation of the 2023 analysis by Georgieva, Ripollés \& McFee.
They found rap to be an outlier with larger F0  variation compared to other genres, but with a declining trend since the genre's inception. 
However, they only analyzed data through 2010. 
Our analysis looks beyond 2010. 
We once again observe rap's large F0 variation, but with a decelerated decline in recent years. 
\end{abstract}
\section{Introduction and Motivation}\label{sec:introduction}


Rap is a genre with highly dynamic rhythmic and vocal styles \cite{ohriner2019flow}.
While the genre is known for its evoking grooves that complement the syncopated verbal expression of the rapper \cite{Coscarello2003wordout}, rap also features rich variations in the fundamental vocal frequency, or F0. 

Recently, Georgieva et al. \cite{Georgieva} carried out a study to quantify the degree of F0 variation across musical genres, rap included. 
They used a dataset with 43,000 songs released between 1924 and 2010. 
Using the F0 total variation (TV) metric \cite{panteli2017towards}, they found that rap showcased larger vocal F0 TV compared to all other genres. 
They also found that this value has significantly declined over the years since the genre's inception.  

This late-breaking demo is a continuation of work by Georgieva et al. \cite{Georgieva}, focusing on rap songs that were popular between 2009 and the present day. 
We zero in on rap not only for being a TV outlier, but also because it is a relatively new musical genre. In fact, while the study by Georgieva et al. \cite{Georgieva} spanned almost a century of music, rap only shows up in the most recent 25 years of their analysis. 
Therefore, their observations around rap deserve a closer look and validation beyond 2010. 

\section{Background and related work}

Why is a continuation of the analysis by Georgieva et al. \cite{Georgieva} beyond 2009 needed? Since the early 2000s, the internet has changed how music is produced, exchanged among creators, and shared with their audience  \cite{o2006consuming}. 
The online availability of free music production software newly means that a hit song can be made by anyone, in a few minutes, and at an unprecedentedly low cost \cite{bell2018dawn}. 
In the specific case of rap, these advancements caused a diversification into sub-genres like trap, drill, and mumble rap, among others \cite{bradley2017anthology}.
The internet also brought ``globalized'' views to music creation, including the broad use of Auto-Tune even in genres traditionally rooted outside Europe or the USA \cite{roman2023f0}.
As a consequence of artists leveraging online platforms to promote their music, a rap song nowadays is often understood as a means of cultivating social engagement in addition to being a ``piece of music'' \cite{burns2017hip}. 

Rap also underwent important developments in the 2010s. There are two main reasons.
First, many rappers cultivated their sound with the main goal of optimizing virality. 
This phenomenon can be traced back to Soulja Boy’s ``Crank Dat'' in 2007 \cite{craven2019soulja}, with the most successful example being Lil Nas X’s ``Old Town Road'' in 2018 \cite{coscarelli2019old}.
Second, during the mid-late 2010s ``SoundCloud Rap'' \cite{battan019takeover} marked the genre's global dominance in terms of popularity \cite{thompson2017shazam}. 
``SoundCloud Rap'' features highly reductive and repetitive flows, resulting in "mumble rap". Its vocal delivery usually operated at the extremes of either slurred lethargy or ferocious intensity, and in some cases was so inarticulate as to be considered ``post-verbal'' \cite{waugh2020every}\footnote{We mention this not as as a value judgement of ``SoundCloud Rap'', but as a means of highlighting its emphasis on raw evocative expression over cleverly crafted poeticism.} 
Modern rap is diverse and in no way confined to the styles of the ``SoundCloud Rap'' era, but suffice it to say that it is no longer fair to only evaluate rap according to metrics of lyrical prowess valued by previous generations \cite{Abraham2018problem}.

\section{Dataset} \label{sec:dataset}
Our data curation methodology ensures that each song is culturally relevant. 
We select songs using \textit{XXL} hip-hip magazine's "Freshman Class"\footnote{www.xxlmag.com/every-xxl-freshman-class-brought-to-hip-hop/}, an annual list that, since 2007, represents each year's stylistic trends by featuring the hottest up-and-coming rappers deemed worthy of further exposure by a combination of editors and fans \cite{kinoti2018hip}.

We used the Spotify API to obtain a list of songs released by every rapper during their \textit{XXL} freshman year.
Using each song's popularity metric we ranked an artist's songs in a given year and selected only the three most popular. 
The resulting dataset includes 409 songs spanning from 2009-2023. 
Since the \textit{XXL} Freshman Class list skipped 2008, our dataset starts in 2009 to avoid gaps. 



\section{Methods}\label{Method}

To acquire each song's audio, we used the \texttt{ytmusicapi} and \texttt{pytube} libraries. 
The resulting audio tracks have a sampling rate of 44,100Hz. 
Next, consistent with Georgieva et al. \cite{Georgieva}, we extracted the vocal stems 
using Demucs \cite{defossez2019demucs}. 
To obtain each vocal stem's  F0 contour, we used \texttt{librosa.pyin} \cite{mcfee2023librosa} with parameters \texttt{fmin=70}, \texttt{fmax=900}, \texttt{frame\_length=2048}, \texttt{hop\_length=512}, \texttt{n\_thresholds=5}, and \texttt{no\_trough\_prob=0.99}. 
This resulted in an F0 contour with a temporal resolution of $\sim$86Hz. 
We removed non-voiced sections for a continuous view of the contour and converted it from Hz to cents to work on a perceptually-linear scale. 
Finally, we calculated the F0 TV
\begin{equation}
    TV(\mathbf{x})=\frac{1}{N}\sum_{i=1}^{N-1}|x_{i+1}-x_i|,
\end{equation}
where $\mathbf{x}=(x_1,x_2,...,x_N)$ is a song's F0 contour. 
It measures the average difference between each instantaneous F0 value and the next one. 
After calculating each song's F0 TV, we use linear regression to quantify trends over years.

Rappers differentiate themselves from their peers via particular F0 contour features, such as interval jumps and variations, upward glissandi, and large slopes on final syllables \cite{ohriner2019analysing}. This is to say that F0 TV alone is capable of capturing meaningful trends in rap vocal practice.


\section{Results and discussion}\label{sec:results}

Our analysis is a continuation of the findings by Georgieva et al. \cite{Georgieva} (see Figure 4 in their publication), where they discovered rap to have overall high F0 TV values that were significantly declining between the years of 1986 to 2010. 

\begin{figure}
 \includegraphics[width=\columnwidth]{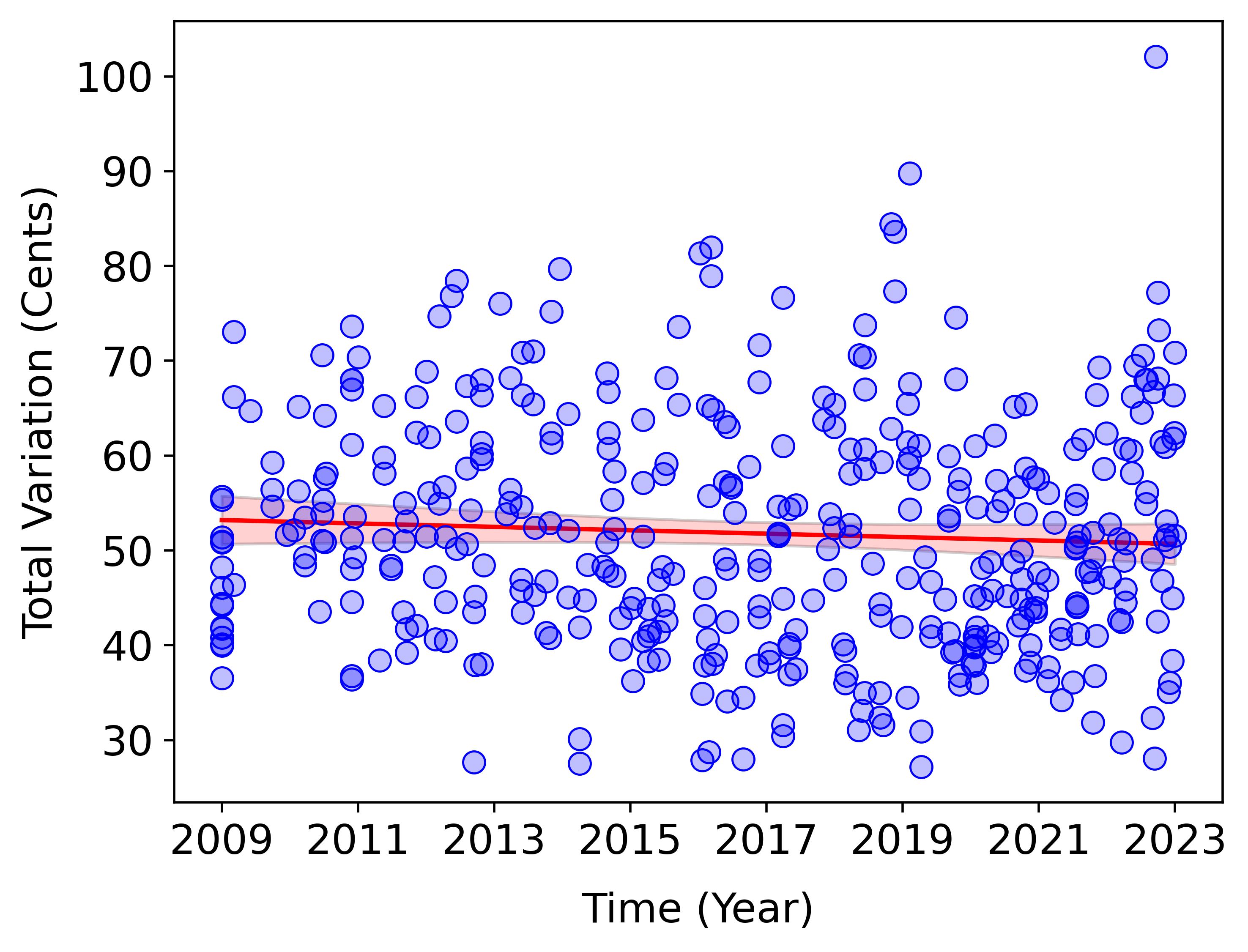}
 \caption{Each dot is a song's vocal F0 TV. Songs were selected based on relevance for a given year's stylistic trends in rap. Linear regression shows a decline over time ($R=-0.061$), but it is not significant ($p=0.218$). This contrasts with Georgieva et al. \cite{Georgieva}, who saw a significant decline in rap vocal F0 TV from 1986-2010. Thus, our study suggests that this trend plateaued in recent years.}
 \label{fig:result}
\end{figure}

Figure \ref{fig:result} shows our linear regression analysis of F0 TV across all songs in our dataset (the shaded red region is the 95\% confidence interval of the regression).
We found rap to have relatively high F0 TV values with a decline ($R=-0.066$; $p=0.181$) from 2009 to 2023.
However, our linear analysis revealed that this decline was not statistically significant beyond 2009 ($p>0.05$). 
In other words, while Georgieva et al. found rap to have a decline in F0 TV from 1986 to 2010 \cite{Georgieva}, our analysis shows that this decline has plateaued since then.

Our analysis also reveals an apparent increase in F0 TV variance between songs for a given year. 
In fact, the standard deviation (SD) of F0 TV values across all songs from 2009 to 2015 is 10.93, while it is 13.11 for songs between 2016 and 2023. 
This trend may be caused by an underlying increase in vocal style variety, a claim consistent with the rise of different sub-genres in the same time period. 

It is worth noting that, compared to Georgieva et al., our study found slightly larger F0 TV values for the year of 2009, which both studies share. 
This difference can be explained by the dataset selection criteria of each study. 
We selected songs that were considered to be ``essential'' and indicative of rap's ``cutting-edge'' for a given year. This selectivity came with the tradeoff of a relatively low number of songs. 
Georgieva et al., in contrast, had a much more comprehensive scope, capturing trends across multiple genres and over longer timespans. As a result, their research analyzed far more data than ours. 


\section{Conclusion}

Our study is a continuation of the original analysis by Georgieva et al. \cite{Georgieva}, who observed that rap has had a decline in F0 TV between the 1980s and the first decade of the 2000s.
We curated a dataset with emblematic rap songs released between the years of 2009 and 2023 to calculate each song's vocal F0 TV. 
Our analysis showed that the decline in F0 TV has decelerated and has perhaps even plateaued in the past 14 years. 
We also observed a general increase in the variability of F0 TV values between songs in a given year, perhaps tied to the increase in the variety of rap styles, but further research is needed to better characterize this trend. 
Our website features the code to reproduce our analysis, as well as listening examples \texttt{iiviiiii.github.io/rap\_f0\_analysis}

\section{acknowledgements}
We thank Andri Mardiansyah \& Smith Tripornkanokrat for help prototyping code for an earlier version of this study. 


\bibliography{ISMIR2023_lbd}

\end{document}